\documentclass{article}
\usepackage{spconf,amsmath,epsfig}

\usepackage[T1]{fontenc}
\usepackage{mathtools}
\usepackage{amsmath}
\usepackage{amsthm}
\usepackage{amssymb}
\usepackage{graphicx}

\usepackage{epstopdf}

\usepackage{multirow}
\usepackage{pgfplots}
\usepackage{algorithmic}
\usepackage{algorithm}
\usepackage{subfigure}
\usepackage{enumitem}
\usepackage{textcomp}

\begin{document}\sloppy

\def\x{{\mathbf x}}
\def\L{{\cal L}}

\title{GENERALIZED RESIDUAL VECTOR QUANTIZATION FOR LARGE SCALE DATA}
%
\name{Shicong Liu, Junru Shao, Hongtao Lu*\thanks{*Corresponding author} \thanks{This paper is supported by NSFC (No. 61272247, 61533012, 61472075), the 863 National High Technology Research and Development Program of China ( SS2015AA020501) and the Major Basic Research Program of Shanghai Science and Technology Committee (15JC1400103).} \thanks{
		978-1-4799-7082-7/15/\$ 31.00 \textcopyright 2015 IEEE}}
\address{\{artheru, yz\_sjr, htlu\}@sjtu.edu.cn \\
	Key Laboratory of Shanghai Education Commission for \\  Intelligent Interaction and Cognitive Engineering, \\
	Department of Computer Science and Engineering,\\ Shanghai Jiao Tong University, P.R.China}
%
%
%

\maketitle

\begin{abstract}
	\small
	Vector quantization is an essential tool for tasks involving large scale data, for example, large scale similarity search, which is crucial for content-based information retrieval and analysis. In this paper, we propose a novel vector quantization framework that iteratively minimizes quantization error. First, we provide a detailed review on a relevant vector quantization method named \textit{residual vector quantization} (RVQ). Next, we propose \textit{generalized residual vector quantization} (GRVQ) to further improve over RVQ. Many vector quantization methods can be viewed as the special cases of our proposed framework. We evaluate GRVQ on several large scale benchmark datasets for large scale search, classification and object retrieval. We compared GRVQ with existing methods in detail. Extensive experiments demonstrate our GRVQ framework substantially outperforms existing methods in term of quantization accuracy and computation efficiency.
\end{abstract}
\section{Introduction}
\label{sec:intro}

With the rapid development of data collecting and mining techniques,
there is an urgent need for powerful algorithms in data compression,
storage and retrieval. Specifically,
compressing a high-dimensional vector and performing similarity search without decompression on large scale data have become
crucial in many fields,
e.g. object detection  \cite{vedaldi2012sparse}, image and video retrieval
\cite{sivic2003video}, and deep neural networks \cite{rodger2014fuzzy} etc.

Vector quantization (VQ) based methods, e.g. \emph{product quantization} (PQ)
\cite{pq}, \emph{optimized product quantization} (OPQ)  \cite{opq}, \emph{additive quantization} (AQ), 
\emph{composite quantization} (CQ)  \cite{composite}, are popular
and successful methods for the tasks above. Vector quantization is essentially lossy compression of high-dimensional vectors. It compresses a vector into a short encoding representation by multiple learned codebooks, and approximately reconstructs the vector by codewords corresponding to the encoding.
Quantization-based algorithms have three major advantages: (1) Memory consumption is significantly
reduced to represent high-dimensional vectors; (2) It allows efficient similarity computation, e.g, one can compute Euclidean distance or scalar products between an uncompressed vector and a large set of compressed vectors via \textit{asymmetric distance computation} (ADC)
\cite{pq} or its variants like \textit{optimized asymmetric distance}  \cite{wang2014optimized}
, hence approximate nearest neighbor
search (ANN) can be greatly accelerated; (3) These encodings are simple
enough to allow more sophisticated data structure and heuristic non-exhaustive
search scheme like \emph{inverted file system with
	asymmetric distance computation} (IVFADC)  \cite{pq}, \emph{inverted
	multi index}  \cite{babenko2012inverted} and \emph{locally optimized
	product quantization}  \cite{kalantidis2014locally}. They are capable of
storing one billion compressed vectors in memory and conducting a retrieval
in a few milliseconds even on a modern laptop. 

In this paper, we propose \textit{generalized residual vector quantization} (GRVQ) to further improve over existing vector quantization methods. The main idea is to iteratively select a codebook and optimize it with the current residual vectors, then re-quantize the dataset to obtain the new residual vectors for the next iteration. GRVQ shares a similar motivation with the traditional \textit{residual vector quantization} (RVQ) ( \cite{rvq},  \cite{vec}). RVQ uses \textit{additive model} to quantize vectors, and adopts a multi-stage residual clustering scheme to learn codebooks. However RVQ fails to generate effective encodings for high-dimensional data \cite{wei2014projected}, which manifests as the information entropy of the encodings obtained on each adding stage drops quickly. Compared to RVQ, our GRVQ: 

\textbf{Overcomes the downsides of RVQ} with \textit{transition clustering} which substantially improves performance of k-means on high intrinsic dimensional data. We also propose a \textit{multi-path encoding} scheme to further lower the quantization error.

\textbf{Generalizes RVQ} that RVQ can be viewed as a special case of GRVQ performing codebook optimization on an \textit{"all-zero"} codebook on every stage.

Compared to the existing vector quantization methods, GRVQ has the following merits:
\begin{enumerate}[leftmargin=*]
	\item Existing vector quantization methods working on \textit{additive model} generally require an extra fix ( \cite{babenko2014additive},  \cite{rvq}) for Euclidean distance computation. GRVQ can eliminate this extra fix by introducing regularization on the codebook learning phase.
	\item Quantizing a vector with \textit{additive model} is NP-hard. Though many approaches have been proposed, e.g, iterated conditional modes \cite{lafferty2001conditional} and AQ-encoding  \cite{babenko2014additive}, they're too slow for practical application. The codebooks obtained by GRVQ are variance descending, enabling a much more efficient and practical beam search encoding scheme.
\end{enumerate}

The quantization accuracy and computation efficiency of our method is validated on three large scale dataset commonly used for evaluating vector quantization methods. We also demonstrate the superior performance of GRVQ on classification task.

\section{Related Methods}
\subsection{Vector Quantization}
Vector quantization (VQ) techniques are used to perform lossy compression on a large scale dataset.
Denote a \emph{database} $\mathcal{X}$ as a set of $N$ $d$-dimensional
vectors for VQ to compress, VQ learns a \emph{codebook} $\mathbf{C}$, which is a list of $K$ \textit{codewords}: $\mathbf{c}(1)$, $\mathbf{c}(2)$, $\ldots$, $\mathbf{c}(K) \in \mathbb{R}^d$. Then VQ uses a \emph{mapping function} $i(\cdot)\,:\,\mathbb{R}^{d}\,\to\,[K]$\footnote{$[K]$ denotes $\{1,\,2,\,3,\,\ldots,\,K\}$} to encode a vector: $\mathbf{x}\mapsto i(\mathbf{x})$. \emph{Quantizer} $q$ is defined as $q(\mathbf{x})=\mathbf{c}(i(\mathbf{x}))$,
meaning $\mathbf{x}$ is approximated as $q(\mathbf{x})$ for latter use. Vector quantization minimizes \emph{quantization error}, which is defined as
\begin{equation}
	E=\frac{1}{N}\sum_{\mathbf{x}\in\mathcal{X}}\lVert\mathbf{x}-q(\mathbf{x})\rVert^{2}.\label{E}
\end{equation}

Minimizing Eqn.\ref{E} directly
leads to classical k-means clustering algorithm  \cite{kmeans}.
VQ essentially partitions the data space into many \emph{Voronoi
	cells}, and quantizes vectors to the \emph{centroids} of the cells. The k-means model is simple and intuitive. However, the cost of training and storing the centers grows linearly with $K$, limiting quantization accuracy. 

\subsection{Residual Vector Quantization}\label{exx}

With a compositional model, one can represent cluster centers more efficiently. A number of compositional models are proposed, e.g, \textit{product quantization} (PQ) \cite{pq},  \emph{optimized product quantization} (OPQ) \cite{opq}, \emph{additive quantization} (AQ) \cite{babenko2014additive}, \emph{composite quantization} (CQ) \cite{composite}. Here we focus on residual vector quantization (RVQ) ( \cite{rvq},  \cite{juang1982multiple}) for high dimensional data lossy compression. RVQ is a common technique to approximate original data with several low complexity quantizers, instead of a prohibitive high complexity quantizer. RVQ algorithm
iteratively learns $M$ quantizers step by step. In the $m$-th step,
RVQ obtains the current residuals $\left\{ \mathbf{e}_\mathbf{x}= \mathbf{x}-\sum_{i=1}^{m-1}q_{i}(\mathbf{x})\,:\,\mathbf{x}\in\mathcal{X}\right\} $
with the previous learned quantizers $\{q_{i}\,:\,1\le i<m\}$. Next, it performs classical k-means to
learn the $m$-th quantizer for the following objective:

\vspace{-0.9\baselineskip}
\begin{equation}
	\min_{\mathbf{c}_{m}(\cdot),\,i_{m}(\cdot)} \frac{1}{N} \sum _{\mathbf{x}\in \mathcal{X}}\lVert\mathbf{e}_\mathbf{x}-\mathbf{c}_{m}(i_{m}(\mathbf{x}))\rVert^{2}.
\end{equation}

\vspace{-0.4\baselineskip}The original vectors are quantized with the following \textit{additive model}:

\vspace{-1.05\baselineskip}
\begin{equation}
	q(\mathbf{x})=\sum_{m=1}^M\mathbf{c}_m(i_m(\mathbf{x}))
\end{equation}

The above \textit{additive model} is also used in AQ, CQ, \textit{tree quantization} \cite{babenko2015tree}, etc. Such model is beneficial for applications like high-dimensional nearest neighbor retrieval, for example, \textit{asymmetric distance computation} (ADC)  \cite{pq}  \cite{babenko2014additive} allows exhaustive nearest neighbor search by efficiently computing Euclidean distance between an uncompressed query vector $\mathbf{q}$ and the compressed dataset vectors $\mathbf{x} \in \mathcal{X}$ with the following equation:
\begin{equation}
	\label{epsilonFix}
	\begin{split}
		\lVert\mathbf{q}-&\mathbf{x}\rVert^2 \approx  \lVert \mathbf{q}- q(\mathbf{x}) \rVert^2 \\
		& = \sum_{m=1}^M\lVert \mathbf{q}-\mathbf{c}_m(i_m(\mathbf{x})))\rVert^2-(m-1)\lVert \mathbf{q}\rVert^2 + \epsilon\\
		\text{where} &\quad \epsilon = \sum_{a=1}^M\sum_{b=1,b\neq a}^M  {\mathbf{c}_a(i_a(\mathbf{x}))}^\mathrm{T}\mathbf{c}_b(i_b(\mathbf{x}))
	\end{split}
\end{equation}

To retrieve the nearest neighbors, we first compute and store $\lVert\mathbf{q}-\mathbf{c}_{m}(k)\rVert^{2}$ for $k\in[K]$ and $m\in[M]$ in a look-up table. As for term $\epsilon$, it can be computed during the dataset compression and stored along with the compressed dataset. Then, the approximate distance between $\mathbf{q}$ and any dataset vector $\mathbf{x}$ can be efficiently computed with $M$ floating point addition. CQ regularizes $\epsilon$ to a fixed value to further reduce the cost for storing the compressed dataset and the cost for computing the approximate distance, \textit{sparse composite quantization}  \cite{zhang2015sparse} proposes a method to accelerate the look-up table computation.

\vspace{-0.5\baselineskip}
\begin{figure}[htbp]
	\begin{center}
		\subfigure[PQ]{
			\includegraphics[width=0.17\linewidth]{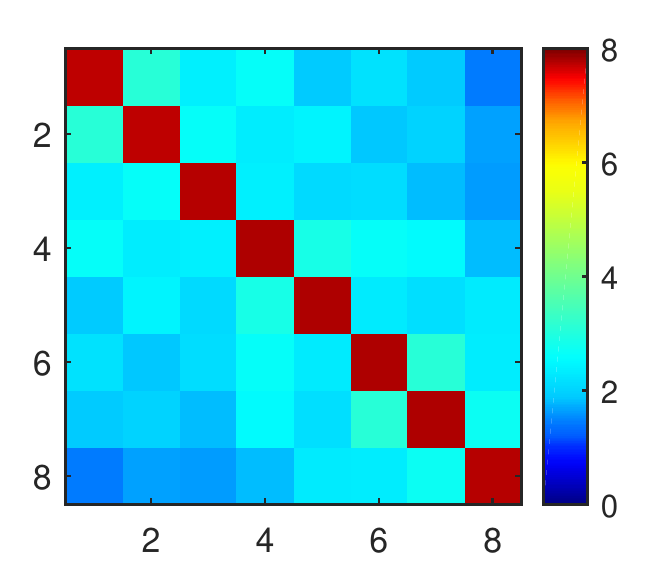}
		}
		\subfigure[RVQ]{
			\includegraphics[width=0.17\linewidth]{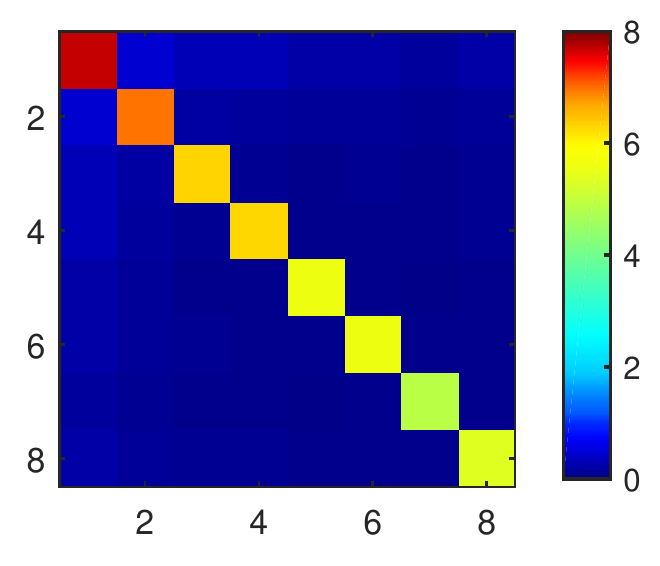}
		}
		\subfigure[GRVQ]{
			\includegraphics[width=0.165\linewidth]{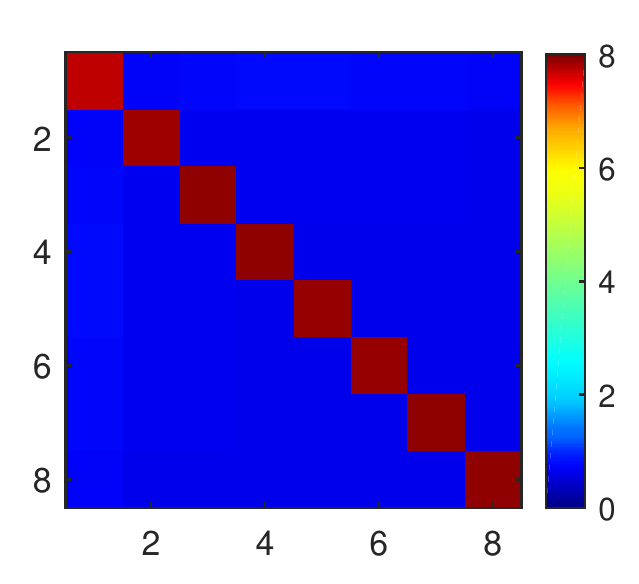}
		}           
		\subfigure[AQ]{
			\includegraphics[width=0.163\linewidth]{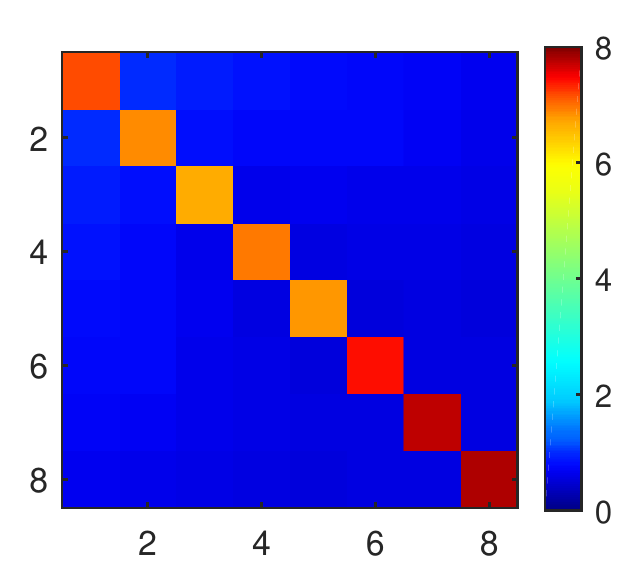}
		}
		\subfigure[OPQ]{
			\includegraphics[width=0.168\linewidth]{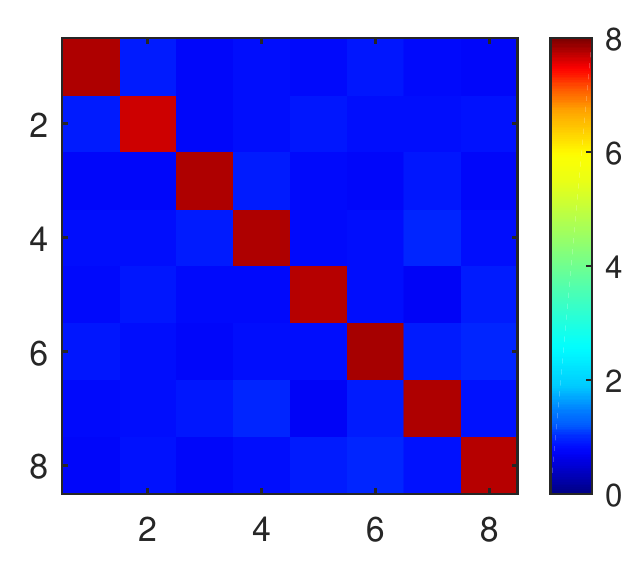}
		}           
		\vspace{-0.5\baselineskip}
		\caption{\small\textit{Mutual information matrices for different quantization methods}. Experiment conducted on GIST-1M dataset. We learned $M=8$ codebooks, $K=256$ codewords per codebook. The perfect encoding should have no mutual information between different codebooks and an information entropy of $\log K=8$-bits for each codebook. Our proposed GRVQ achieves near optimal encoding.}
		\label{figmm}
	\end{center}
\end{figure}

\subsection{Disadvantages of Residual Vector Quantization}
RVQ quantized vectors have relatively higher quantization errors compared to other vector quantization methods in high dimensional space, as observed in Fig. \ref{ds}, the performance gain of adding an additional stage drops quickly for RVQ. 

We examine the encodings obtained by vector quantization from the view of information entropy. For an effective encoding, the information entropy of encoding at any position should be high, and the mutual information of encoding at different positions should be low. Note the above objective is explicitly considered in Hashing methods like \textit{spectral hashing}  \cite{SH}. To formulate, denote the encoding at position $m$ as a discrete random variable $I_m$ with domain of $[K]$, and $H(\cdot)$ as the information entropy of a random variable, we would like following equations to hold:

\vspace{-1.5\baselineskip}
\begin{equation}
	\begin{split}
		H(I_{m}) & \approx\log_{2}K, \quad s.t. \quad m\in[M]\\
		H(I_{i};\,I_{j}) & \approx 0, \quad s.t. \quad  i,\,j\in[M], i\ne j;\\
	\end{split}
\end{equation}

\vspace{-0.5\baselineskip}
RVQ doesn't produce encodings with $H(I_{m})$ high enough, as observed in Fig. \ref{figmm}. This is mainly because the intrinsic dimensionality of the residual vectors becomes higher with increasing stages  \cite{wei2014projected}, hence traditional k-means algorithm fails to work. AQ  \cite{babenko2014additive} has a slightly higher $H(I_{m})$ compared to RVQ, yet still much lower than other quantization methods like OPQ. Thus an improvement over this should be beneficial.

In addition, as an additive model, the quantization of a vector is actually a fully connected discrete MRF problem ( \cite{composite},  \cite{babenko2014additive}). However RVQ doesn't consider this in the codebook learning. Thus on each stage, the codebook is not learned with the optimal input. This leads to an accumulating quantization error and impact the overall quantization accuracy.

\section{Generalized Residual Vector Quantization}
\begin{algorithm}[t]
	\small
	\caption{Generalized Residual Vector Quantization}
	\label{algIOlearn}
	\textbf{Input}: $d$-dimensional dataset $\mathcal{X}$, containing $N$ vectors; number of codebooks $M$; number of elements $K$ per codebook; initial codebooks $\mathbf{C}_m=\{\mathbf{c}_m(1), \cdots, \mathbf{c}_m(K) \}, m\in[M]$.
	
	\textbf{Output}: Optimized codebooks: $\{\mathbf{C}_m : m\in[M]\}$
	
	\begin{algorithmic}[1]
		\REPEAT 
		
		\STATE Encode $\mathbf{x}\in\mathcal{X}$ with \textit{multi-path encoding} in Sec.\ref{sec:multi-path-encoding}, and obtain the residual $\mathbf{e}_{\mathbf{x}}$ defined in Eqn. \ref{ex}.
		\STATE Randomly pick a codebook $\mathbf{C}_m$, generate an intermediate dataset $\mathcal{X}'$ defined in Eqn. \ref{xprime}.
		
		\STATE Optimize $\mathbf{C}_m$ for Eqn. \ref{keq}, with \textit{transition clustering} algorithm described in Sec. \ref{sec:improved-clustering-algorithm}.
		
		\UNTIL {Quit Condition}
	\end{algorithmic}
\end{algorithm}

We propose \emph{generalized residual vector quantization} (GRVQ) to learn effective encodings with additive model. We present the outline of GRVQ in Algorithm \ref{algIOlearn}. GRVQ optimizes existing codebooks or codebooks of \textit{zero vectors} to learn from scratch. Formally, denote the encoding of $\mathbf{x} \mapsto (i_1(\mathbf{x}), i_2(\mathbf{x}), \ldots, i_m(\mathbf{x}))$, the current residual of $\mathbf{x}$ is:

\vspace{-0.7\baselineskip}
\begin{equation}
	\mathbf{e}_{\mathbf{x}} =\mathbf{x}-\sum_{m=1}^{M}\mathbf{c}_{m}(i_{m}(\mathbf{x})).
	\label{ex}
\end{equation}
On each iteration, GRVQ randomly pick an $m$-th codebook $\mathbf{C}_{m}=\{\mathbf{c}_{m}(1),\cdots,\mathbf{c}_{m}(K)\}$ to optimize. We first perform incremental clustering on an intermediate dataset $\mathcal{X}'$, defined as:
\begin{equation}
	\mathcal{X}^{\prime}=\left\{ \mathbf{x}^{\prime} | \mathbf{x}^{\prime}=\mathbf{e}_{\mathbf{x}}+\mathbf{c}_{m}(i_{m}(\mathbf{x})), \mathbf{x}\in\mathcal{X} \right\}.
	\label{xprime}
\end{equation}

Then we optimize the codebook to fit this dataset better with the following objective function:
\begin{equation}
	\min_{\mathbf{c}_{m}(\cdot),\,i(\cdot)}\frac{1}{N}\sum_{\mathbf{x}^\prime\in\mathcal{X}^{\prime}}\lVert\mathbf{x}^{\prime}-\mathbf{c}_{m}(i(\mathbf{x}^{\prime}))\rVert^{2}.
	\label{keq}
\end{equation}

Finally, we re-encode the original dataset $\mathcal{X}$ with the optimized codebooks $\mathbf{C}_1, \cdots,\mathbf{C}_K$, and obtain the residual vectors for the next iteration. 

Traditional RVQ is a special case of GRVQ which is initialized on \textit{all zeros} codebooks, and sequentially optimizes each codebook for once. \textit{Product quantization}  \cite{pq} and \textit{optimized product quantization} can be viewed as GRVQ with constraints that each codebook only works on specific dimensions. 

\subsection{Transition Clustering}\label{sec:improved-clustering-algorithm}

The increased randomness of high-dimensional residual vectors lead to the failure of traditional k-means algorithm  \cite{wei2014projected}. In order to obtain a better clustering result, a typical approach is to cluster on lower-dimensional subspace  \cite{agrawal1998automatic}, with the objective function:
\begin{equation}
	\small
	\min_{\mathbf{c}(\cdot),\,i(\cdot), R}\frac{1}{N}\sum_{\mathbf{x}\in\mathcal{X}^{\prime}}\lVert R\mathbf{x}-\mathbf{c}(i(R\mathbf{x}))\rVert^{2} \quad s.t. \quad R^{\top}R=I_{k}
\end{equation}

Note that k-means algorithm on the entire $d$-dimensional space
is a special case when $\mathrm{rank}\left(R\right)=d$. \textit{Transition clustering} seek a
\emph{transition} from subspace clustering to the full dimensional
clustering. we first use PCA dimension reduced subspace to initialize the clustering  \cite{ding2004k}, then iteratively add more dimensions and warm start k-means algorithm with the clustering information obtained from previous iteration, as they provide good starting position \cite{bradley1998refining}. To optimize a codebook $\mathbf{C}_{m}$ for Eqn.\ref{keq}, we perform the following:
\begin{enumerate}[leftmargin=*]
	\setlength\itemsep{0.1em}
	
	\item Designate a dimension increasing sequence: $d_{1}<d_{2}<\cdots<d_{I}=d$\footnote{We choose parameters $I=10$ and $d_{i}=d^{i/I}$ in our experiments.} ;
	\item Project $\mathbf{C}_{m}$ and $\mathcal{X}'$ into PCA space $R$
	of $\mathcal{X}^{\prime}$: $\mathbf{C}_{m}^{r}=\{\mathbf{c}_{m}^{r}(k)=R\mathbf{c}_{m}(k)\,:\, k\in [K]\}$, and  $\mathcal{X}^{r}=\{\mathbf{x}^{r}=R\mathbf{x}^{\prime}\,:\mathbf{x}^{\prime}\in\mathcal{X}^{\prime}\}$;
	\item Perform warm-started k-means initialized with the first $d_{i}$ dimensions of $\mathbf{C}_{m}^{r}$, and update them with the resulting centroids. We do this iteratively for $i=1\cdots I$.
	\item Rotate $\mathbf{C}_{m}^{r}$ back to finish the optimization:
	$\mathbf{C}_{m}=\{R^{\top}\mathbf{c}_{m}^{r}(k)\,:\,1\le k\le K\}$.
\end{enumerate}

\subsection{Multi-path Encoding}\label{sec:multi-path-encoding}

\label{encoding}

Encoding with additive model is a fully-connected MRF problem (\cite{babenko2014additive},  \cite{composite}). Though it can be solved approximately by various existing
algorithms, they are very time consuming  \cite{babenko2015tree}. In this section we propose an efficient beam search method for GRVQ optimized codebooks.

Denote $\mathbf{x}\mapsto (i_1, i_2, \cdots, i_M) $ as the optimal encodings for $\mathbf{x}$, which quantizes $\mathbf{x}\approx\sum_{m=1}^M\mathbf{c}_{m}(i_{m})$ minimizing the quantization error $E=\lVert\mathbf{x}-\sum_{m=1}^{M}\mathbf{c}_{m}(i_{m}(\mathbf{x}))\rVert^{2}$.
Suppose we know the first $(n-1)$ optimal encodings $(i_{1},\,i_{2},\,\cdots,\,i_{n-1})$. To determine the $n$-th optimal encoding $i_n$ effectively,
denote $\hat{\mathbf{x}}=\sum_{m=1}^{n-1}\mathbf{c}_{m}(i_{m})$
and $\widetilde{\mathbf{x}}=\sum_{m=n+1}^{M}\mathbf{c}_{m}(i_{m})$, and consider quantization error $E$ as a function of $i_n$:
\begin{equation}
	\begin{split}E= & \lVert\mathbf{x}-(\hat{\mathbf{x}}+\mathbf{c}_{n}(i_{n})+\widetilde{\mathbf{x}})\rVert^{2}\\
		=&\lVert\mathbf{x}-\hat{\mathbf{x}}-\widetilde{\mathbf{x}}\rVert^2+\lVert \mathbf{c}_{n}(i_{n}) \rVert^2 \\
		& -2\mathbf{c}_{n}(i_{n})^{\top} (\mathbf{x}-\hat{\mathbf{x}})- 2\mathbf{c}_{n}(i_{n})^{\top}\widetilde{\mathbf{x}}
	\end{split}
	\label{equ2}
\end{equation}

We seek the best $i_{n}$ in $[K]$, in order to minimize
$E$. In Eqn. \ref{equ2}, term $2\mathbf{c}_{n}(i_{n}))^{\top}\widetilde{\mathbf{x}}$
cannot be computed because $\widetilde{\mathbf{x}}$ is unknown to the
encoding scheme, which leads to an error in estimating the best $i_n$. 
Low variance of $\widetilde{\mathbf{x}}$ is required for neglecting $2\mathbf{c}_{n}(i_{n}))^{\top}\widetilde{\mathbf{x}}$.
A simple way to achieve this goal is to rearrange the codebooks by the variance of codewords in descending order. In fact, GRVQ naturally produces codebooks descending order of variance of corresponding codewords.

Then, we can we adopt the idea behind the beam search algorithm to encode a vector
$\mathbf{x}$. That is, we sequentially encode $\mathbf{x}$ with each codebook and maintain a list of $L$ best encodings of $\mathbf{x}$. On each iteration, we enumerate all possible codewords on the next codebook, compute the distortion, and determine the new $L$ best encodings. This can be done efficiently with lookup tables. The time complexity of encoding with an $m$-th codebook is $O(dK+mKL+KL\log L)$. To encode a vector with all $M$ codebooks, the time complexity is $O(dMK+M^2KL+MKL\log L)$. 

One should notice that when GRVQ has optimized an $m$-th codebook, there is no need to re-encode the vectors with the first $(m-1)$ codebooks since our method is carried out sequentially and will obtain exactly the same first $(m-1)$ encodings. This is very different from the encoding scheme proposed in  \cite{babenko2014additive}, in which the change in any codebook requires re-encoding over all codebooks. Our encoding scheme is also much more efficient compared to  \cite{babenko2014additive}, because we are only required to consider one codebook at a time.

With the encoding time vs quantization error curve presented in Fig. \ref{ee},
we find that $L=10$ already could achieve good encoding
quality and have relatively low encoding time. We use this configuration in the rest of the experiment.

\subsection{Eliminating $\epsilon$-term for Efficient Euclidean Distance Computation}

\begin{table}
	\centering
	\setlength{\tabcolsep}{0.3em}
	\footnotesize
	\begin{tabular}{|c|c|c|c|}
		\hline  & Elimination & Quantization & Don't care \\ 
		\hline Processing Time & Long & Short & No \\ 
		\hline Quantization Error & High & Low & Low \\ 
		\hline Extra Length & No & 6-8 bit & 4 Byte \\ 
		\hline Computation & No & 2 Flops & 1 Flops \\ 
		\hline 
	\end{tabular}

	\caption{\small Comparison of different methods for processing $\epsilon$}\label{tbeps}
\end{table}

Quantization methods using additive model like AQ and RVQ require an extra $\epsilon$-term fix to perform Euclidean ADC as mentioned in Sec.\ref{exx}. \textit{Composite quantization} is a method similar to AQ, only that it eliminates this $\epsilon$-term fix by imposing regularization that for all vectors $\epsilon=\epsilon_0$ is a constant.

Similarly, we can modify the objective functions for \textit{transition clustering} and \textit{multi-path encoding} to eliminate $\epsilon$. We introduce regularization parameter $\lambda$ indicating the penalty, and a target parameter $\epsilon_0$. We modify Eqn.\ref{keq} to the following:
\begin{equation}
	\min_{\mathbf{c}_{m}(\cdot),\,i_m(\cdot)}\frac{1}{N}\sum_{\mathbf{x}^\prime\in\mathcal{X}^{\prime}}(\lVert\mathbf{x}^{\prime}-\mathbf{c}_{m}(i_m(\mathbf{x}^{\prime}))\rVert^{2}+\lambda (\epsilon-\epsilon_0)^2)
	\label{keq2}
\end{equation}

The above problem can be solved via a slight modification to k-means algorithm employed in transition clustering: on each iteration of k-means, we assign a vector $\mathbf{x}$ to the centroid that minimize $\lVert\mathbf{x}-\mathbf{c}(i(\mathbf{x}))\rVert^{2}+\lambda (\epsilon-\epsilon_0)^2$.

Next, on encoding a vector $\mathbf{x}$, we optimize the following equation to satisfy the $\epsilon$ regularization:

\vspace{-0.8\baselineskip}
\begin{equation}
	\min_{i_m(\cdot)} \lVert \mathbf{x}-\sum_{m=1}^{M}\mathbf{c}_m(i_m(\mathbf{x})) \rVert+\lambda (\epsilon-\epsilon_0)^2
\end{equation}

The above problem only requires a trivia modification to \textit{multi-path encoding} simply by considering this penalty in the beam search. 

We start the elimination with $\lambda=0$, then on each iteration of GRVQ we compute $\epsilon_0=\mathtt{mean}(\epsilon)$, and slightly increase $\lambda$ to enforce the regularization. The regularization on $\epsilon$ put a slight loss on quantization accuracy as observed in Fig.\ref{ee} for accelerated distance computation and lower memory consumption. Another option is to quantize $\epsilon$ into a few bit \cite{babenko2014additive}, we don't need long code for quantizing $\epsilon$ as observed in Fig.\ref{qe}. We compare different ways of processing $\epsilon$ in Table \ref{tbeps}.
\subsection{Extensibility for online codebook learning}

\begin{figure*}
	\centering
	\begin{center}
		\subfigure[]{
			\includegraphics[width=0.130\linewidth]{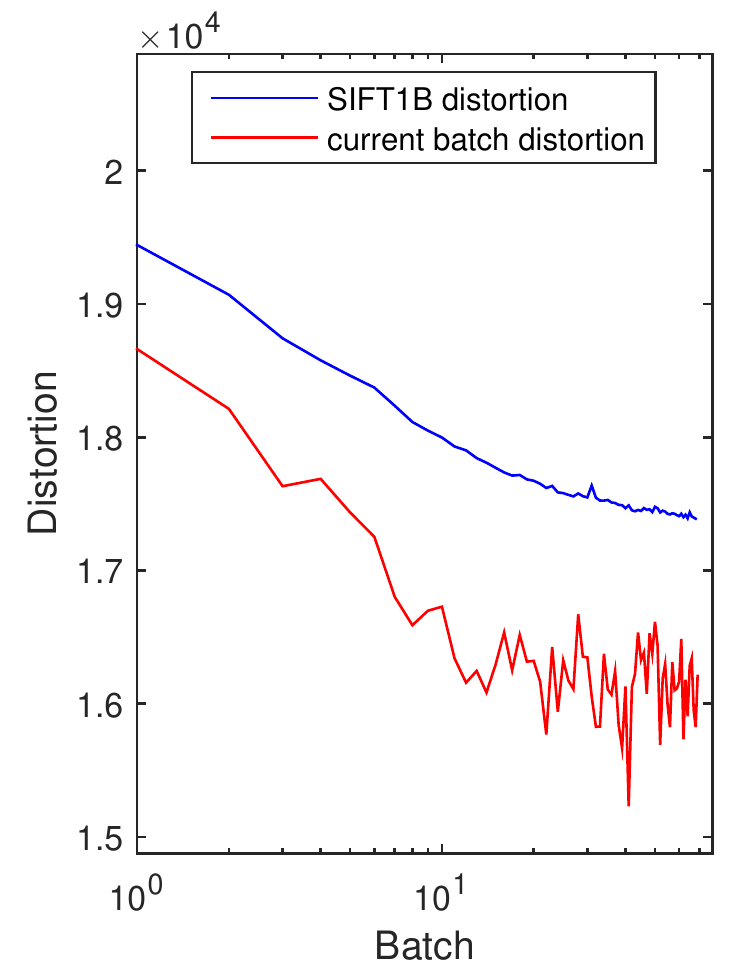}
			\label{online}
		}
		\subfigure[]{
			\includegraphics[width=0.165\linewidth]{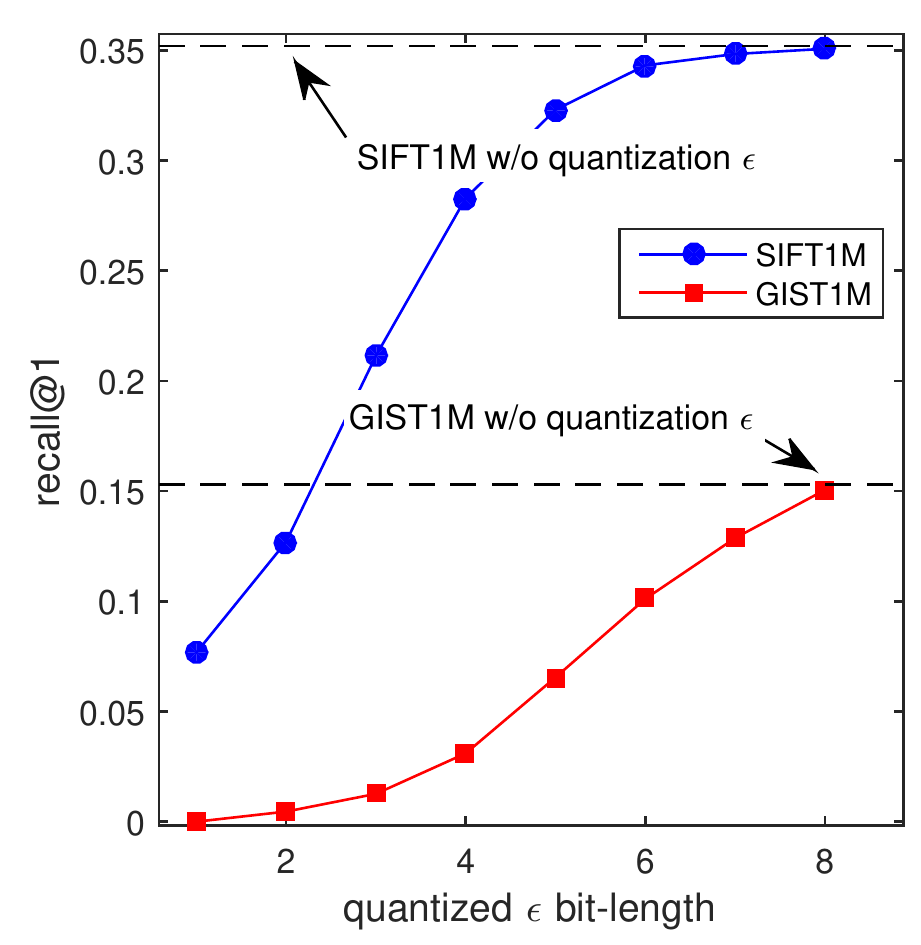}
			\label{qe}
		}
		\subfigure[]{
			\includegraphics[width=0.150\linewidth]{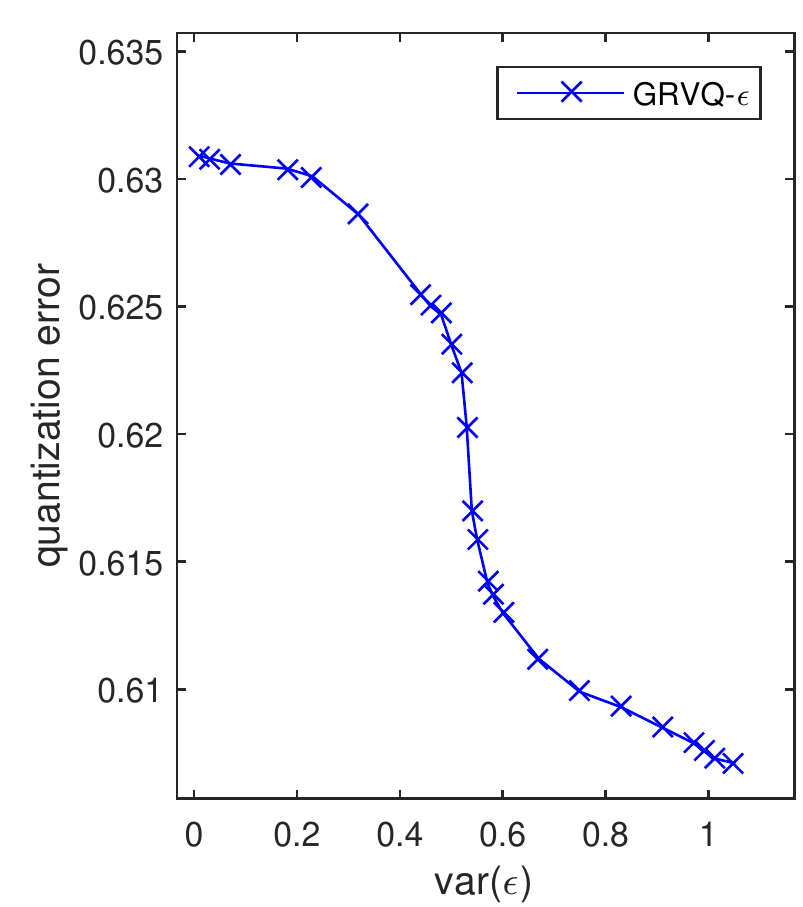}
			\label{ee}
		}           
		\subfigure[]{
			\includegraphics[width=0.157\linewidth]{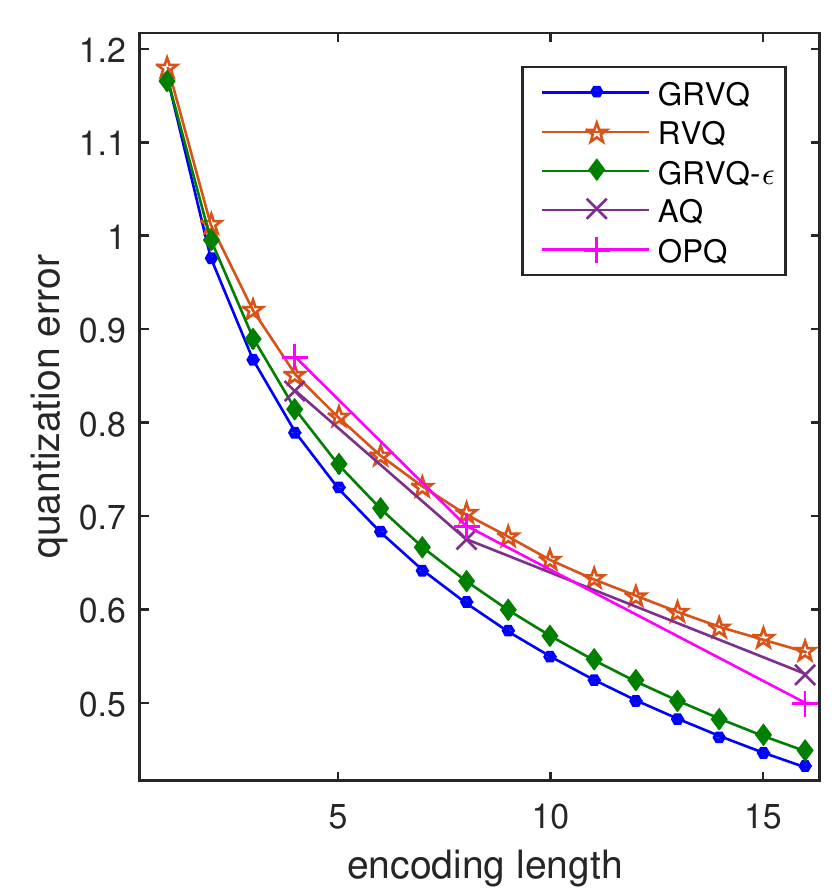}
			\label{ds}
		}     
		\subfigure[]{
			\includegraphics[width=0.15\linewidth]{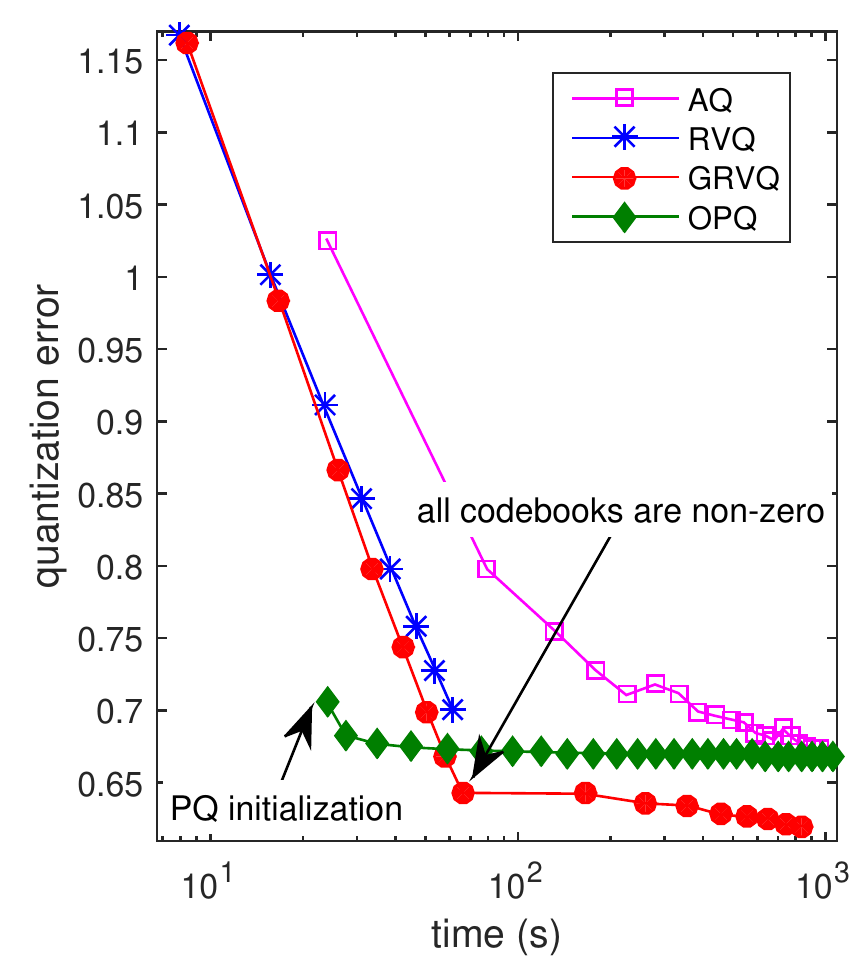}
			\label{time}
		}   
		\subfigure[]{
			\includegraphics[width=0.165\linewidth]{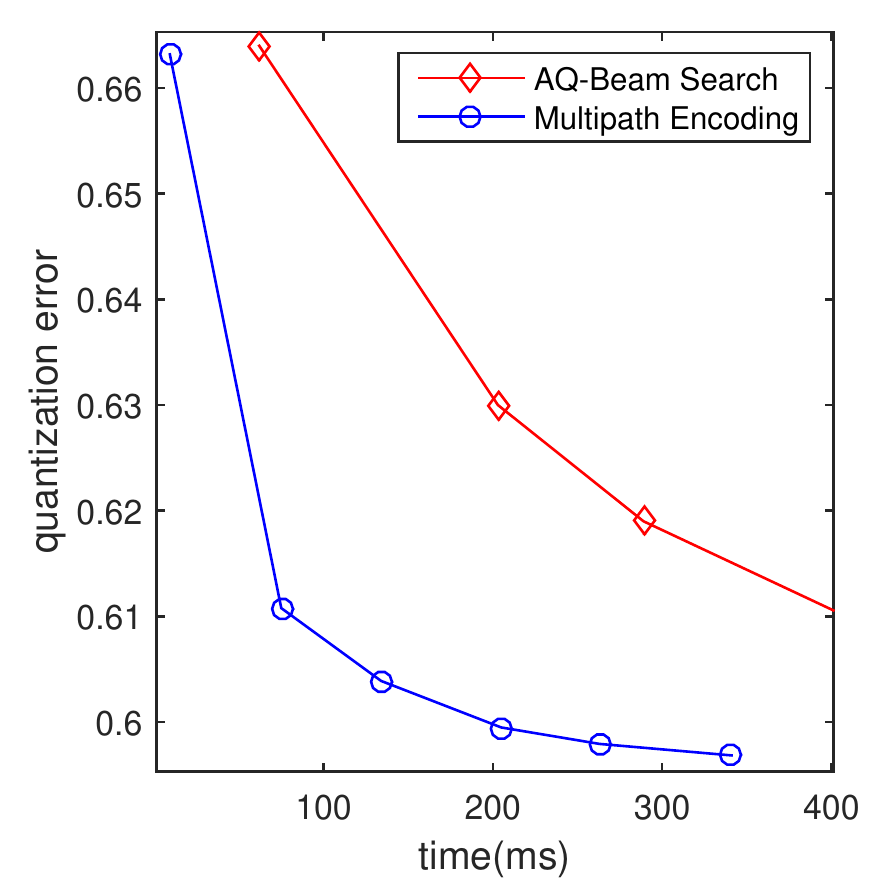}
			\label{enc}
		}   
		\caption{\small All analysis are based on $M=8, K=256$ codebooks. \textbf{(a)} \emph{Online learning effect on SIFT1B} \textbf{(b)} \textit{Quantization effect on $\epsilon$} \textbf{(c)} \textit{Optimization curve of $\epsilon$ elimination on GIST1M, initialized with GRVQ $M=8, K=256$ codebooks.} \textbf{(d)} \textit{Encoding length vs quantization error on GIST1M} \textbf{(e)} \textit{Training time vs quantization error on GIST1M} \textbf{(f)} \textit{Encoding speed comparison with additive codebooks for GIST1M}
		}
		\label{figm}
	\end{center}
\end{figure*}

GRVQ is naturally an online learning mechanism that is able to deal with incrementally obtained training data. This can be done simply by optimizing the codebooks on the new-coming data.
It is also capable of handling large scale dataset where classical clustering algorithm is prohibitive due to unacceptable time complexity and memory consumption. Online learning effect on \textbf{SIFT1B}  \cite{jegou2011searching} dataset containing one billion vectors is reported in
Figure \ref{online}.

\section{Experiments}

\subsection{Dataset and configurations}
In this section we present the experimental evaluation of GRVQ. All experiments are done on a quad-core CPU running at 3.5GHz with 16G memory and one GTX980 GPU.

We use the following datasets commonly used for evaluating vector quantization methods: \textbf{SIFT1M} \cite{pq}, contains one million 128-d
SIFT  \cite{sift} features. \textbf{GIST1M} \cite{pq}, contains one million 960-d
GIST  \cite{gist} global descriptors. \textbf{SIFT1B} \cite{jegou2011searching} contains one billion 128-d SIFT feature as base vectors.

We compare GRVQ with the following state-of-the-art VQ methods:
PQ  \cite{pq}, OPQ \cite{opq}, AQ \cite{babenko2014additive}. We choose the commonly used configuration for codebooks learning: $K=256$, and $M=8,16$. We train all methods on the training set and encode the base dataset. We train online version of GRVQ with all the data. The training time on GIST1M of all methods is presented on Fig.\ref{time}. Though PQ / OPQ train fast, the performance gets easily saturated. Performance of AQ appears unstable. The proposed GRVQ achieves a balanced trade-off between performance and training speed. We draw the time for encoding $1000$ vectors from GIST1M with GRVQ learned $M=8, K=256$ codebooks in Fig.\ref{enc}. Our proposed multi-path encoding scheme utilizes the characteristic of GRVQ codebooks and encodes efficiently. Fig.\ref{ds} shows our GRVQ outperforms existing methods by large margin on all code length ranging from $M=1\cdots16$ in term of quantization accuracy.

\subsection{Large Scale Search}
\begin{figure}
	\includegraphics[width=1\linewidth]{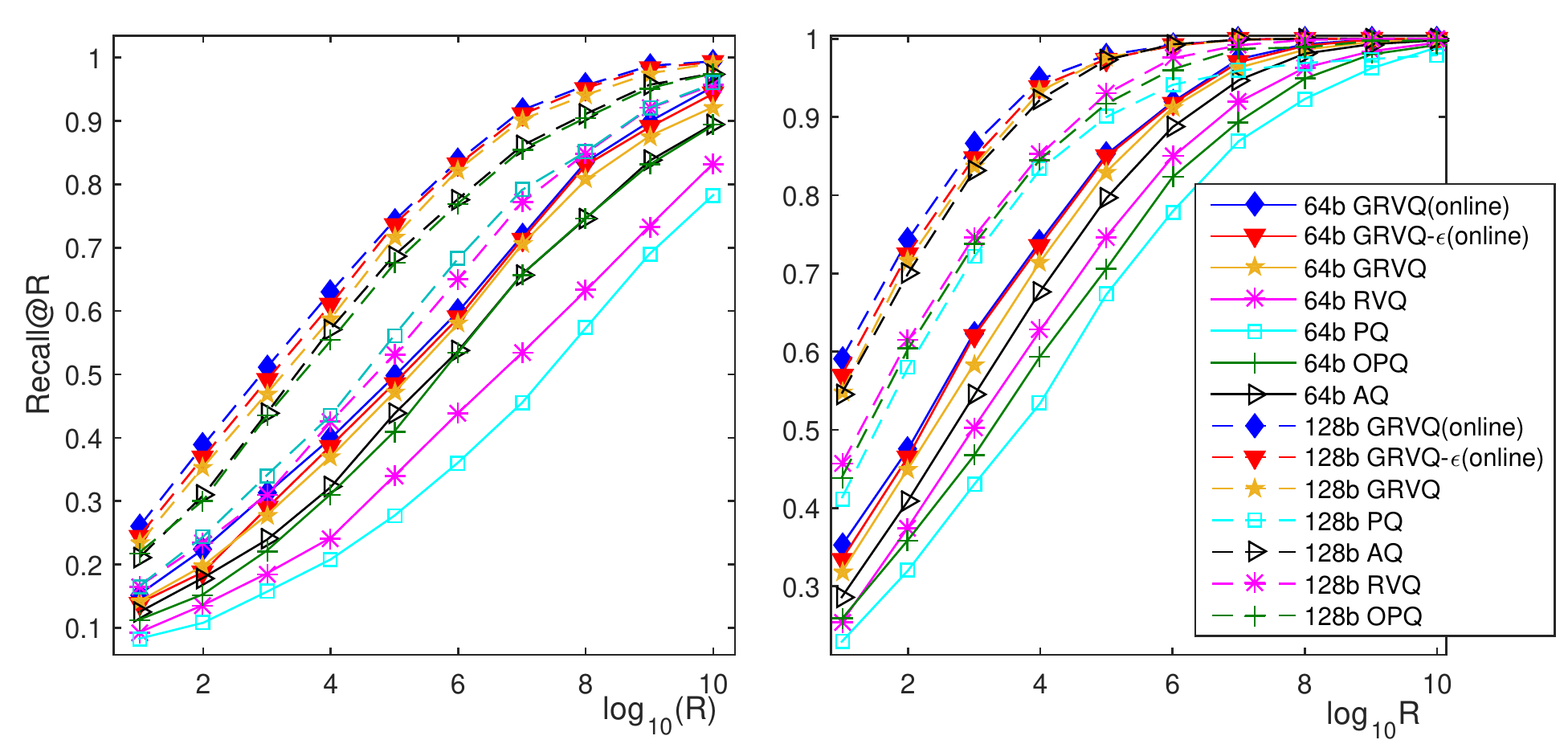}
	\caption{\small Performance for different algorithms for searching the true nearest neighbors. \textit{Left}: performance on \textbf{GIST1M}, \textit{Right}: performance on \textbf{SIFT1M}}
	\label{perf}
\end{figure}

We perform exhaustive Euclidean nearest neighbor search to compare different vector quantization methods. Fig.\ref{perf} shows the results of large scale datasets SIFT1M and GIST1M.  It can be seen that the gains obtained by our approaches are significant on both datasets. The online version of our GRVQ outperforms existing methods by large margin, for example, the performance of 64-bit GRVQ encoding closely match the performance of 128-bit PQ encoding. The $\epsilon$-term eliminated GRVQ codebooks also achieves large improvement over other methods. 

Table \ref{1b} shows the performance for an even larger dataset SIFT1B. By utilizing online learning, our method achieves the best performance. The improvement on large dataset is more significant than that on smaller datasets.

\begin{table}
	\centering \footnotesize
	\setlength{\tabcolsep}{0.3em}
	\begin{tabular}{|c|c|c|c|c|}
		\hline  & GRVQ(online) & CQ & OPQ & PQ \\ 
		\hline Recall@100 & 0.834 & \textit{(0.701)} & \textit{(0.646)} & 0.581 \\ 
		\hline 
		
	\end{tabular} 
	\caption{\small The performance of NN-search on SIFT1B dataset in terms of Recall$@$100. Data in parentheses are taken from  \cite{composite}.}
	\label{1b}
\end{table}

\subsection{Image Classification and Retrieval}
Another important application of vector quantization is to compress image descriptors for image classification and retrieval, in which images are usually represented as the aggregation of local descriptors, result in vectors of thousand dimensions. We evaluate the image classification and retrieval performances over $d=4096$ dimensional \textit{fisher vectors} \cite{perronnin2010improving} on 64-d PCA dimension reduced SIFT descriptors extracted form \textit{INRIA holiday dataset} \cite{jegou2008hamming}. We quantize all fisher vectors and learn a linear classifier to perform the classification. With short codes one can accelerate the classification and retrieval thousand-fold. In particular, our GRVQ has the minimal degradation of performance.
\begin{table}
	\setlength{\tabcolsep}{0.3em}
	\centering \footnotesize
	\begin{tabular}{|c|c|c|c|c|c|c|}
		\hline
		&  GRVQ  &   AQ   &  OPQ   &  RVQ   &   PQ   &  CQ \\ \hline
		32bit & \textbf{57.1}\% & 54.5\% & 53.7\% & 50.9\% & 50.3\% &  (\textit{55.0\%}) \\ \hline
		64bit & \textbf{62.9}\% & 62.1\% & 57.9\% & 53.8\% & 55.0\% &  (\textit{62.2\%}) \\ \hline
	\end{tabular} 
	\caption{\footnotesize The performance of classification over INRIA holiday in terms of mAP. We use compressed vectors for learning linear classifier, and uncompressed vectors for classification. The classification mAP without vector quantization is 63.3\%. Data in parentheses are taken from  \cite{composite}. }
\end{table}

\section{Conclusion}
In this paper, we proposed the \textit{generalized residue vector quantization} (GRVQ) to perform vector quantization with higher quantization accuracy. We proposed \textit{improved clustering algorithm} and \textit{multi-path encoding} for GRVQ codebook learning and encoding. We also propose $\epsilon$-free version of GRVQ for efficient Euclidean distance computation. Experiments against several state-of-the-art quantization methods on well known datasets demonstrate the effectiveness of GRVQ on a number of applications.

\bibliographystyle{IEEEbib}
\footnotesize

\bibliography{camera-ready_icme2016template}

\end{document}